\documentstyle[amssymb,preprint,aps,epsfig]{revtex}

\begin{document}
\title{Analogy between a two-well Bose-Einstein condensate and atom diffraction}
\author{H. L. Haroutyunyan and G. Nienhuis}
\address{Huygens Laboratorium, Universiteit Leiden,\\
Postbus 9504, \\
2300 RA Leiden, The Netherlands}
\maketitle

\begin{abstract}
We compare the dynamics of a Bose-Einstein condensate in two coupled
potential wells with atoms diffracting from a standing light wave. The
corresponding Hamiltonians have an identical appearance, but with a
different set of commutation rules. Well-known diffraction phenomena as {\it %
Pendell\H{o}sung} oscillations between opposite momenta in the case of Bragg
diffraction, and adiabatic transitions between momentum states are shown to
have analogies in the two-well case. They represent the collective exchange
of a fixed number of atoms between the wells.
\end{abstract}


\section{Introduction}

The most common approach to the description of a trapped Bose-condensed gas
is based on the mean-field approximation, which yields the Gross-Pitaevski
equation for the macroscopic wave function. This wave function, which
depends on the number of atoms, plays the role of the mode function for the
Maxwell field. This approach is reliable when the condensate is trapped in a
single quantum state in a potential well. However, when the condensate is
separated into two or more parts, so that more than one quantum state is
populated, the mean-field approach is not evidently justified. It has been
shown by Javanainen and Yoo \cite{Javanainen} that two originally separate
parts of a condensate that are initially in a Fock state and that are
brought to overlap will reveal an interference pattern that varies in
position from one realization to another. This effect, which has also been
observed experimentally \cite{Ketterle} cannot be described by a single
macroscopic wavefunction. A simple model for a condensate in a double
potential well is defined by a field-theoretical Hamiltonian for a
boson-Hubbard dimer \cite{Kalosakas,Ivanov}, which can be expressed in terms
of SU(2) angular momentum-type operators with a quadratic term. This latter
term represents the interaction between atoms in a well. The mean-field
approximation is basically equivalent to classical equations of motion for
the expectation values of the SU(2) operators \cite{Milburn,Anglin}. The
quantum regime has mainly been studied numerically, leading to collapse and
revival \cite{Milburn}, and to non-classical dynamics arising from the
periodic modulation of the coupling between the wells \cite{Salmond}. The
formation of a two-well condensate by the raising of the barrier has been
analyzed theoretically \cite{Menotti}. The situation of a Bose-Einstein
condensate (BEC) in a two-well trap is also studied experimentally \cite
{Thomas,Tiecke}.

A very similar Hamiltonian describes the situation of an atom diffracting
from a standing-wave optical potential. This problem has received attention
already in the early days of laser cooling \cite{Pritchard2}. More recent
work has developed the band structure of the energy spectrum \cite
{Schumacher}, and a number of regimes have been distinguished that allow an
analytical description \cite{Keller}. In a simple version of the model, the
Hamiltonian is identical in form as in the two-well problem mentioned above.
Now the quadratic term represent the kinetic energy of the atom. The only
difference between the two cases is that the commutation rules for the
operators in the diffraction case are slightly simplified compared to the
case of SU(2) symmetry.

In this paper we discuss the analogy and the differences between these two
systems. We point that a number of analytical solutions known for the
diffraction problem can be carried over to the two-well system. The physics
of these cases is discussed.

\section{BEC in a double potential well}

We consider a potential consisting of two wells. When the barrier between
the wells is not too low, the ground state and the first excited state $%
\left| g\right\rangle $ and $\left| e\right\rangle $ of a single atom are
well approximated as the even and odd superposition of the lowest bound
states in the two wells. Therefore, these states can be described as 
\begin{equation}
\left| g\right\rangle =\frac{1}{\sqrt{2}}(\left| 1\right\rangle +\left|
2\right\rangle )\;;\;\left| e\right\rangle =\frac{1}{\sqrt{2}}(\left|
1\right\rangle -\left| 2\right\rangle )\;,  \label{bec1}
\end{equation}
with $\left| 1\right\rangle $ and $\left| 2\right\rangle $ the localized
states in either well. When the energy separation between the excited and
the ground state is indicated as $\hbar \delta $, the off-diagonal element
of the one-particle Hamiltonian $\widehat{H}_{1}$ between the localized
states is 
\[
\left\langle 1\right| \widehat{H}_{1}\left| 2\right\rangle =-\hbar \delta
/2\;. 
\]
At the low energies that are of interest here, the two particle interaction
is well approximated by the standard contact potential $U({\vec{r}},{\vec{r}}%
^{\prime })=(4\pi \hbar ^{2}a/m)\delta ({\vec{r}}-{\vec{r}}^{\prime })$,
with $a$ the scattering length. The second-quantized field operator is now 
\begin{equation}
\widehat{\Psi }({\vec{r}})=\widehat{a}_{g}\psi _{g}({\vec{r}})+\widehat{a}%
_{e}\psi _{e}({\vec{r}})=\widehat{a}_{1}\psi _{1}({\vec{r}})+\widehat{a}%
_{2}\psi _{2}({\vec{r}})\;,  \label{Psi}
\end{equation}
in terms of the wavefunctions $\psi _{i}$ and the annihilation operators $%
\widehat{a}_{i}$ of the single-particle states. The annihilation operators
and the corresponding creation operators obey the standard bosonic
commutation rules. The corresponding Hamiltonian is 
\begin{equation}
\widehat{H}=\int d{\vec{r}}\,\widehat{\Psi }^{\dagger }({\vec{r}})H_{1}%
\widehat{\Psi }({\vec{r}})+\int d{\vec{r}}\,d{\vec{r}}^{\prime }\,\widehat{%
\Psi }^{\dagger }({\vec{r}})\widehat{\Psi }^{\dagger }({\vec{r}}^{\prime })U(%
{\vec{r}},{\vec{r}}^{\prime })\widehat{\Psi }({\vec{r}})\widehat{\Psi }({%
\vec{r}}^{\prime })\;.  \label{Hamfield}
\end{equation}
The wavefunctions $\psi _{1}$ and $\psi _{2}$ of the localized states have
the same form, and we assume that they do not overlap. Then the interaction
term can be expressed exclusively in the parameter $\kappa $ defined by 
\begin{equation}
\hbar \kappa ={\frac{4\pi \hbar ^{2}a}{m}}\int d{\vec{r}}\;|\psi _{1}({\vec{r%
}})|^{4}\;,  \label{kappa}
\end{equation}
which measures the strength of the interatomic interaction. Performing the
integrations in eq. (\ref{Hamfield}) leads to the expression for the
Hamiltonian 
\begin{equation}
\widehat{H}=-\frac{\hbar \delta }{2}\left( \widehat{a}_{1}^{\dagger }%
\widehat{a}_{2}+\widehat{a}_{2}^{\dagger }\widehat{a}_{1}\right) +\frac{%
\hbar \kappa }{2}\left( \widehat{a}_{1}^{\dagger }\widehat{a}_{1}^{\dagger }%
\widehat{a}_{1}\widehat{a}_{1}+\widehat{a}_{2}^{\dagger }\widehat{a}%
_{2}^{\dagger }\widehat{a}_{2}\widehat{a}_{2}\right) \;,  \label{bec3}
\end{equation}
where we took the zero of energy halfway the two energy levels of a single
atom. This is also known as the boson-Hubbard dimer Hamiltonian \cite
{Kalosakas}.

The Hamiltonian (\ref{bec3}) can also be expressed in terms of SU(2)
operators by applying the standard Schwinger representation of two modes.
This leads to the definition 
\begin{equation}
\widehat{J}_{0}={\frac{1}{2}}\left( \widehat{a}_{1}^{\dagger }\widehat{a}%
_{1}-\widehat{a}_{2}^{\dagger }\widehat{a}_{2}\right) \;,\;\widehat{J}_{+}=%
\widehat{a}_{1}^{\dagger }\widehat{a}_{2}\;,\;\widehat{J}_{-}=\widehat{a}%
_{2}^{\dagger }\widehat{a}_{1}\;.  \label{bec5}
\end{equation}
These operators are related to the Cartesian components of angular momentum
by the standard relations $\widehat{J}_{\pm }=\widehat{J}_{x}\pm i\widehat{J}%
_{y}$, and $\widehat{J}_{0}=\widehat{J}_{z}$. They obey the commutation
rules for angular momentum operators 
\begin{equation}
\lbrack \widehat{J}_{0},\widehat{J}_{\pm }]=\pm \widehat{J}_{\pm }\;,\;[%
\widehat{J}_{+},\widehat{J}_{-}]=2\widehat{J}_{0}\;,  \label{bec6}
\end{equation}
which generate the SU(2) algebra. The Hamiltonian (\ref{bec3}) can be
rewritten in the form 
\begin{equation}
\widehat{H}=-\frac{\hbar \delta }{2}(\widehat{J}_{+}+\widehat{J}_{-})+\hbar
\kappa \widehat{J}_{0}^{2}+\frac{\hbar \kappa }{4}\left( \widehat{N}^{2}-2%
\widehat{N}\right) \;,  \label{Hamop}
\end{equation}
with $\widehat{N}=\widehat{a}_{1}^{\dagger }\widehat{a}_{1}+\widehat{a}%
_{2}^{\dagger }\widehat{a}_{2}$ the operator for the total number of
particles. Obviously, the Hamiltonian (\ref{Hamop}) commutes with $\widehat{N%
}$, and it is block-diagonal in the number of particles $N$. For each value
of $N$, the Hamiltonian (\ref{Hamop}) can be expressed as 
\[
\widehat{H}_{N}+\frac{\hbar \kappa }{4}(N^{2}-2N)\;, 
\]
with the $N$-particle Hamiltonian 
\begin{equation}
\widehat{H}_{N}=-\frac{\hbar \delta }{2}(\widehat{J}_{+}+\widehat{J}%
_{-})+\hbar \kappa \widehat{J}_{0}^{2}\;,  \label{HN}
\end{equation}
where the operators are now restricted to the $N+1$ Fock states $\left|
n,N-n\right\rangle $ with $n=0,1,\dots N$, with $n$ particles in well $1$,
and $N-n$ particles in well $2$. In the language of angular momentum, this
manifold of states corresponds to the angular-momentum quantum number $J=N/2$%
, and the $2J+1$ Fock states are eigenstates of $\widehat{J}_{0}$ with
eigenvalue $\mu =n-N/2$, with $\mu =-J,-J+1,\dots ,J$. Note that $\mu $ is
half the difference of the particle number in the two wells. For an even
number of particles, the angular-momentum quantum number $J$ as well as the
'magnetic' quantum numbers are integer, whereas these number are
half-integer in case of an odd number of particles. The action of the
operators $\widehat{J}_{0}$ and $\widehat{J}_{\pm }$ on the Fock states has
the well-known behavior 
\begin{equation}
\widehat{J}_{0}\left| \mu \right\rangle =\mu \left| \mu \right\rangle \;,\;%
\widehat{J}_{+}\left| \mu \right\rangle =f_{\mu +1}^{\text{ }}\left| \mu
+1\right\rangle ,\widehat{J}_{-}\left| \mu \right\rangle =f_{\mu }^{\text{ }%
}\left| \mu -1\right\rangle  \label{Jpm}
\end{equation}
with $f_{\mu }^{\text{ }}=\sqrt{\left( J+\mu \right) \left( J-\mu +1\right) }
$. The $\mu $-dependence of the strength of the hopping operators $\widehat{J%
}_{\pm }$ reflects the bosonic accumulation factor, which favors the arrival
of an additional bosonic atom in an already occupied state.

When the quadratic term in eq. (\ref{HN}) would be replaced by a linear
term, the evolution would be a uniform rotation in the $2J+1$-dimensional
state space with angular frequency $\sqrt{\delta^2 + \kappa^2}$. The
presence of the quadratic term makes the dynamics considerably more complex.
Therefore we compare this dynamics with another well-known case in which a
similar quadratic term appears.

\section{Standing-wave diffraction of atoms}

The translational motion of a two-level atom in a far detuned standing-wave
light field is described by the effective Hamiltonian 
\begin{equation}
\widehat{H}_{d}=-\frac{\hbar ^{2}}{2m}\frac{\partial ^{2}}{\partial z^{2}}-%
\frac{\hbar \omega _{R}{}^{2}}{\Delta }\text{ }\cos ^{2}kz\;,
\label{lattice2}
\end{equation}
with $\Delta =\omega _{0}-\omega $ the difference of the resonance frequency
and the optical frequency, and $\omega _{R}$ the Rabi frequency of each of
the travelling waves that make up the standing wave. The Hamiltonian takes a
particularly simple form in momentum representation, since the
kinetic-energy term is diagonal in momentum, and the potential energy
changes the momentum by $\pm 2\hbar k$. Therefore, we introduce momentum
eigenstates $|\mu \rangle $ which have the momentum $2\mu \hbar k$. Then
apart from an irrelevant constant, the Hamiltonian (\ref{lattice2}) can be
represented in the algebraic form 
\begin{equation}
\widehat{H}_{d}=-\frac{\hbar \delta }{2}\left( \widehat{B}_{+}+\widehat{B}%
_{-}\right) +\hbar \kappa \widehat{B}_{0}^{2}\;,  \label{lattice12}
\end{equation}
where $\kappa =2\hbar k^{2}/m$ determines the kinetic energy term, and $%
\delta =\omega _{R}^{2}/2\Delta $ the atom-field coupling. The operators
occurring on the r.h.s. are defined by the relations 
\begin{equation}
\widehat{B}_{0}|\mu \rangle =\mu |\mu \rangle ;\text{ }\widehat{B}_{\pm
}|\mu \rangle =|\mu \pm 1\rangle \;.  \label{lattice4}
\end{equation}
They differ from the corresponding relations (\ref{Jpm}) in that now the
strength of the hopping operators is uniform.

This Hamiltonian (\ref{lattice12}) has the same form as eq. (\ref{HN}), even
though they describe completely different physical situations. The
difference is mathematically characterized by the commutation relations. The
SU(2) relations (\ref{bec6}) are replaced by the simpler set 
\begin{equation}
\lbrack \widehat{B}_{0},\widehat{B}_{\pm }]=\pm \text{ }\widehat{B}_{\pm
}\;,\;[\widehat{B}_{+},\widehat{B}_{-}]=0\;,  \label{lattice6}
\end{equation}
which is easily found from their explicit expressions (\ref{lattice4}). The
two operators $\widehat{B}_{\pm }$ are found to commute. A result of this
difference is that the state space in the two-well case has a finite
dimension $2J+1=N+1$, whereas the momentum space has an infinite number of
dimensions.

A mathematically identical set of operators occurs in the description of the
dynamics of the Wannier-Stark system, consisting of a particle in a periodic
potential with an additional uniform force \cite{article}. In that case, the
eigenstates of $\widehat{B}_{0}$ represent the spatially localized Wannier
states, rather than the momentum states.

We recall three approximate solutions of the evolution governed by the
Hamiltonian (\ref{lattice12}), which are valid in different situations, and
which allow analytical solutions.

The Raman-Nath regime is valid for interaction times that are so short that
the atom has no time to propagate. Then the quadratic term in (\ref
{lattice12}) can be neglected, and the evolution is determined by the
atom-field coupling $\delta (t)$. The evolution operator is simply $\widehat{%
U}=\exp [i\phi (\widehat{B}_{+}+\widehat{B}_{-})/2]$, where $\phi =\int
dt\delta (t)$ is the integral of the coupling constant over the evolution
period. The matrix elements of the resulting evolution operator for the
pulse can be found by operator algebra in the form \cite{article} 
\begin{equation}
\langle \mu ^{\prime }|\widehat{U}|\mu \rangle =i^{\mu ^{\prime }-\mu
}J_{\mu ^{\prime }-\mu }(\phi )  \label{evol}
\end{equation}
in terms of Bessel functions. For an initial state $\left| \mu \right\rangle 
$ with a well-determined momentum, the time-dependent state following the
pulse can be expressed as 
\begin{equation}
\left| \Psi (t)\right\rangle \simeq \sum_{\mu ^{\prime }}e^{-i\kappa t\mu
^{\prime 2}}\left| \mu ^{\prime }\right\rangle \left\langle \mu ^{\prime
}\right| \widehat{U}\left| \mu \right\rangle .  \label{timedep}
\end{equation}
This leads to explicit analytical expressions for diffraction experiments 
\cite{Pritchard2}. The probability of transfer of $n$ units of momentum is
proportional to $|J_{n}(\phi )|^{2}$.

The Bragg regime is valid when the coupling $\delta $ between neighboring
momentum states is small compared to the kinetic-energy separation $\approx
2\hbar \kappa \mu $ of the initial state $|\mu \rangle $ from its
neighboring states $\left| \mu +1\right\rangle $ . This initial state leads
to an oscillating time-dependent state between the two states $|\mu \rangle $
and $|-\mu \rangle $ with the same kinetic energy 
\begin{equation}
\left| \Psi \left( t\right) \right\rangle =\cos \frac{\Omega _{\mu }t}{2}%
\text{ }\left| \mu \right\rangle +i\sin \frac{\Omega _{\mu }t}{2}\text{ }%
\left| -\mu \right\rangle \;,  \label{oscill}
\end{equation}
apart from an overall phase factor. This can only occur when the momentum
transfer $2\mu $ (in units of $2\hbar k$) is an integer, which corresponds
precisely to the Bragg condition. The {\em Pendell\"{o}sung} frequency is
given by $\Omega _{\mu }=\delta (\delta /2\kappa )^{2\mu -1}/\left[ \left(
2\mu -1\right) !\right] ^{2}$ \cite{Keller}. This expression is fully
analogous to the effective Rabi frequency for a resonant multiphoton
transition, with non-resonant intermediate states \cite{Shore,Giltner}.

The regime of adiabatic coupling arises for a time-dependent atom-field
coupling $\delta (t)$ that varies sufficiently slowly, so that an initial
energy eigenstate remains an eigenstate. The adiabaticity condition in the
present case reads 
\begin{equation}
\frac{d\delta }{dt}\ll \kappa \delta  \label{lattice9}
\end{equation}
When an atom passes a standing wave with a sufficiently smooth variation of
the intensity, and the Bragg condition is fulfilled, the presence of two
initially degenerate eigenstates $|\pm \mu \rangle $ leads to interference
after the passage, which produces two outgoing beams. Because of the
similarity between the two Hamiltonians (\ref{HN}) and (\ref{lattice12}),
these well-known diffraction cases can be expected to have analogies in the
dynamics of the two-well problem.

\section{Symmetry considerations of generic Hamiltonian}

The Hamiltonians (\ref{HN}) and (\ref{lattice12}) can be represented in the
generic form 
\begin{equation}
\widehat{H}=-\hbar \delta \widehat{L}_{x}+\hbar \kappa \widehat{L}_{z}^{2},
\label{lattice10}
\end{equation}
with $\widehat{L}_{x}=(\widehat{L}_{+}+\widehat{L}_{-})/2$, $\widehat{L}_{z}=%
\widehat{L}_{0}$, where the operators $\widehat{L}_{i}$ represent $\widehat{J%
}_{i}$ or $\widehat{B}_{i}$, depending on the commutation rules and the
corresponding algebra that they obey. In the two-well case, the eigenstates $%
\left| \mu \right\rangle $ of the operator $\widehat{L}_{z}$ represent
number states in the two-well case, with the eigenvalue $\mu $ half the
number difference between the wells. In the diffraction case, the states $%
\left| \mu \right\rangle $ are momentum eigenstates. In this latter case,
the coupling between neighboring momentum states is independent of $\mu $
(eq. (\ref{lattice4})), whereas in the two-well case the $\mu $-dependence
of the hopping operator indicated in eq. (\ref{Jpm}) reflects the bosonic
accumulation effect. A consequence of this is also that the Hamiltonian in
the diffraction case couples an infinite number of states $\left| \mu
\right\rangle $, whereas in the two-well case the number of coupled states
has the finite value $N+1$. In the diffraction case we restrict ourselves to
the situation that the Bragg condition is respected. Therefore, both in the
diffraction case and in the two-well case $\mu $ attains either integer or
half-integer values. The action of $\widehat{L}_{z}$ is the same in both
cases.

The Hamiltonian (\ref{lattice10}) is invariant for inversion of $\mu $. In
order to demonstrate this, we introduce the inversion operator $\widehat{P}$%
, defined by the relation $\widehat{P}|\mu \rangle =|-\mu \rangle $. In the
diffraction case, the operator $\widehat{P}$ corresponds to inversion of
momentum, which does not change the kinetic energy. In the two-well case,
the operator $\widehat{P}$ represents interchanging the particle numbers in
the two wells, which has no effect on the interparticle interaction. The
commutation rules of the inversion operator with the operators $\widehat{L}%
_{i}$ are specified by $\widehat{P}\widehat{L}_{z}\widehat{P}=-\widehat{L}%
_{z}$, $\widehat{P}\widehat{L}_{\pm }\widehat{P}=\widehat{L}_{\mp }$, so
that $\widehat{P}$ inverts $\widehat{L}_{y}$ and $\widehat{L}_{z}$, and
commutes with $\widehat{L}_{x}$. It follows that the Hamiltonian (\ref
{lattice10}) commutes with $\widehat{P}$, so that it is invariant for
inversion of $\mu $. Therefore the Hamiltonian has vanishing matrix elements
between the even and the odd subspaces, which are the eigenspaces of $%
\widehat{P}$ with eigenvalue $1$ and $-1$ respectively. For half-integer $%
\mu $-values, these spaces are spanned by the states 
\begin{equation}
\left| \mu \right\rangle _{+}\equiv \frac{\left| \mu \right\rangle +\left|
-\mu \right\rangle }{\sqrt{2}};\ \left| \mu \right\rangle _{-}\equiv \frac{%
\left| \mu \right\rangle -\left| -\mu \right\rangle }{\sqrt{2}};
\label{oddeven}
\end{equation}
for positive values of $\mu .$ In the case of integer $\mu $-values, the
state $\left| \mu =0\right\rangle $ also belongs to the even subspace. The
even and odd subspace evolve independently from one another. This symmetry
property of $H$ depends on the fact that it is quadratic in the operator $%
\widehat{L}_{z}$.

The action of the quadratic term in the Hamiltonian (\ref{lattice10}) on the
new basis is simply given by the relation $\widehat{L}_{z}^{2}|\mu \rangle
_{\pm }=\mu ^{2}|\mu \rangle _{\pm }$. The action of the coupling term in
the Hamiltonian can be expressed in a general form by introducing
coefficients $F_{\mu }$ for non-negative values of $\mu $. In the case of
the SU(2) algebra, we define $F_{\mu }=f_{\mu }$, whereas in the diffraction
case we simply have $F_{\mu }=1$. The matrix elements of $\widehat{L}_{x}$
can be fully expressed in terms of the coefficients $F_{\mu }$ for positive $%
\mu $. Within the even or the odd subspace, the operator $\widehat{L}_{x}$
has off-diagonal matrix elements only between two states for which the
values of $\mu $ differ by $1$, and we find 
\begin{equation}
_{\pm }\left\langle \mu +1\left| \widehat{L}_{x}\right| \mu \right\rangle
_{\pm }=\frac{1}{2}F_{\mu +1},  \label{offdiag}
\end{equation}
provided that the value of $\mu $ is positive. These matrix elements
coincide with those on the basis of the states $\left| \mu \right\rangle $.
For the state $\left| \mu =0\right\rangle $, which belongs to the even
subspace of a manifold of states with integer $\mu $-values, the matrix
element is 
\begin{equation}
_{+}\left\langle 1\left| \widehat{L}_{x}\right| 0\right\rangle =F_{1}/\sqrt{2%
}.  \label{offdiag1}
\end{equation}
On the other hand, in a manifold of states with half-integer $\mu $-values, $%
\widehat{L}_{x}$ has a single non-zero diagonal element for $\mu =1/2$, that
is given by 
\begin{equation}
_{\pm }\left\langle 1/2\right| \widehat{L}_{x}\left| 1/2\right\rangle _{\pm
}=\pm \text{ }F_{1/2}.  \label{sym15}
\end{equation}
Hence, in the case of half-integer $\mu $-values, the Hamiltonian projected
on the even and the odd subspace differ exclusively in the diagonal matrix
element for $\mu =\frac{1}{2}$, for which we find 
\begin{equation}
_{\pm }\left\langle 1/2\right| \widehat{H}\left| 1/2\right\rangle _{\pm }=%
\frac{\hbar \kappa }{4}\mp \frac{1}{2}\hbar \delta F_{1/2}.  \label{diag1/2}
\end{equation}
For integer values of $\mu $, the Hamiltonian for the odd subspace is
identical to the Hamiltonian for the even subspace with $\mu \succcurlyeq 1$%
. The only difference is that the even subspace also contains the state $%
\left| 0\right\rangle $, which is coupled to the other states by the matrix
element 
\begin{equation}
_{+}\left\langle 1\left| \widehat{H}\right| 0\right\rangle =\left\langle
0\left| \widehat{H}\right| 1\right\rangle _{+}=-\hbar \delta F_{1}/\sqrt{2.}
\label{even}
\end{equation}

In both cases, the difference between the Hamiltonian parts on the even and
odd subspaces are proportional to $\delta $. These differences are
responsible for the energy splitting between the even and the odd energy
eigenstates. Moreover, since these differences in the Hamiltonian parts
occur for low values of $\mu $, we expect that the even-odd energy
splittings get small for large $\mu $-values. This is confirmed by numerical
calculations. In Figs. 1 and 2 we display the energy levels of the
Hamiltonian, for a few values of $\delta /\kappa $, both for the double-well
case (with $N=100$), and for the diffraction case. The energy levels are
found to be alternatingly even and odd, with increasing energy. In the
two-well case, the energy shifts and splittings due to the coupling are
larger for the same value of $\delta /\kappa $ and the same value of $\mu $.
This arises from the factor $F_{\mu }$, which is unity in the diffraction
case, whereas in the two-well case it decreases from $\sim J=N/2$ at $\mu =0$
to zero at $\mu =J$. In fact, the condition for weak coupling is that matrix
elements coupling the states $\left| \mu \right\rangle $ and $\left| \mu
-1\right\rangle $ are small compared with their unperturbed energy
separation. This condition can be expressed as 
\begin{equation}
\lambda _{\mu }=\frac{\delta }{2\kappa }\frac{F_{\mu }}{2\mu -1}<1.
\label{criteria}
\end{equation}
In the two-well case, the lowest energy states start out to be nearly
equidistant up to $\mu $-values where $\lambda _{\mu }$ approaches one.

\section{Pendell\H{o}sung oscillations}

The energy splittings between the even and the odd eigenstates give rise to
time-dependent states that oscillate between the states $\left| \pm \mu
\right\rangle $. In the diffraction case, they correspond to the well-known
Pendell\H{o}sung oscillations in the Bragg regime. Here we show that similar
oscillations can occur for the two-well problem, and we give an analytical
estimation of the oscillation frequencies. For the generic Hamiltonian given
by (\ref{lattice10}), the Bragg condition is fulfilled when the inequality (%
\ref{criteria}) holds.

The energy differences between the even and odd states to lowest order in $%
\lambda _{\mu }$ \smallskip can be found from the effective Hamiltonian for
two degenerate states that are coupled via a number of non-resonant
intermediate states. This situation occurs for the states $\left| \pm \mu
\right\rangle $, with their $2\mu -1$ intermediate states. In this case, the
intermediate states can be eliminated adiabatically, as demonstrated in Sec.
18.7 of ref. \cite{Shore}. The resulting effective Hamiltonian for these two
states $\left| \pm \mu \right\rangle $ has an off-diagonal element that is
the ratio between two products. The numerator contains the product of the
successive $2\mu $ matrix elements $-\hbar \delta F_{\mu ^{\prime }}/2$ of
the Hamiltonian coupling neighboring states, and the denominator is the
product of the $2\mu -1$ unperturbed energy differences of the degenerate
states $\left| \pm \mu \right\rangle $ with the successive intermediate
states. In the diffraction case, this result coincides with the calculation
given in ref. \cite{Schumacher}, which was obtained by diagonalizing a
tridiagonal matrix and keeping only the lowest order in $\delta /\kappa $.

Generalizing this result to the present case of the two states $\left| \pm
\mu \right\rangle $, we find that the effective Hamiltonian has the diagonal
element 
\begin{equation}
\left\langle \pm \mu \right| \widehat{H}_{eff}\left| \pm \mu \right\rangle
=\hbar \kappa \mu ^{2}  \label{diageff}
\end{equation}
and the off-diagonal element 
\begin{equation}
\left\langle \mp \mu \right| \widehat{H}_{eff}\left| \pm \mu \right\rangle
=-\hbar \Omega _{\mu }/2,  \label{offdiageff}
\end{equation}
with $\Omega _{\mu }$ an effective oscillation frequency given by 
\begin{equation}
\Omega _{\mu }=\left( -1\right) ^{2\mu +1}\frac{1}{2^{2\mu -1}}\frac{\delta
^{2\mu }}{\kappa ^{2\mu -1}}\frac{1}{^{\left[ \left( 2\mu -1\right) !\right]
^{2}}}F.  \label{rabi}
\end{equation}
The factor $F$ is just the product of the coefficients $F_{\mu }$
successively coupling the states intermediate between $\left| \mu
\right\rangle $ and $\left| -\mu \right\rangle $. In the diffraction case,
we simply have $F=1$, whereas in the case of SU(2) symmetry, applying to the
double well we find 
\begin{equation}
F=\frac{\left( J+\mu \right) !}{\left( J-\mu \right) !}  \label{FSU2}
\end{equation}
These expressions are valid both for integer and half-integer values of $\mu 
$. The eigenstates of the effective Hamiltonian are the even and odd states,
and the eigenvalue equations are $\widehat{H}_{eff}\left| \mu \right\rangle
_{\pm }=\left( \hbar \kappa \mu ^{2}\mp \hbar \Omega _{\mu }/2\right) \left|
\mu \right\rangle _{\pm }$. For integer values of $\mu $, the frequency $%
\Omega _{\mu }$ is negative, so that the even states $\left| \mu
\right\rangle _{+}$ are shifted upwards, and the odd states are shifted
downwards in energy. The opposite is true for half-integer values of $\mu $.
In both cases, the ground state is even, and the energy eigenstates for
increasing energy are alternatingly even and odd. In view of the results of
the numerical calculation mentioned above, one may expect that this
alternating behavior of the even and odd eigenstates is valid for all finite
values of the ratio $\delta /\kappa $.

For an initial state $\left| \mu \right\rangle $, this effective Hamiltonian
leads to a time-dependent state that is given by (\ref{oscill}), apart from
an irrelevant overall phase factor. This shows that the oscillating solution
(\ref{oscill}) corresponding to the Bragg regime of diffraction can be
generalized to the case of a condensate in a double well. The same
expression (\ref{oscill}) remains valid, while the oscillation frequency $%
\Omega _{\mu }$ is determined by eqs. (\ref{rabi}) and (\ref{FSU2}). This
describes a state of the condensate atoms in the double well in the
weak-coupling limit. In this case, the state oscillates between the Fock
states $\left| n_{1},n_{2}\right\rangle =\left| N/2+\mu ,N/2-\mu
\right\rangle $ and $\left| n_{1},n_{2}\right\rangle =\left| N/2-\mu
,N/2+\mu \right\rangle $.

Obviously, when the initial state is given by $\left| \mu \right\rangle
_{\pm }$, the system is in a stationary state, and no oscillations occur. In
this case, Pendell\H{o}sung oscillations can still be induced by including
in the Hamiltonian a term that is linear in $\widehat{L}_{z}$. In the
diffraction case, there is no obvious physical realization of such a term.
For the Wannier-Stark system, where the quadratic term in $\widehat{L}%
_{z}^{2}$ is absent, the linear term can be realized by imposing a uniform
force, which gives rise to Bloch oscillations \cite{article,Salomon}. In the
case of the BEC in a double well a term $\hbar \xi \widehat{L}_{z}/2$ in the
Hamiltonian can be realized by imposing an energy difference $\hbar \xi $
between the single-particle ground states in the two wells. When this term
is periodically varying, it can be used for coherent control of the
condensate \cite{Holthaus}. The additional term couples the even and odd
subspaces, thereby breaking the symmetry opf the Hamiltonian. On the basis
of the states $\left| \mu \right\rangle _{\pm }$ the effective Hamiltonian
attains the off-diagonal element 
\begin{equation}
_{\pm }\left\langle \mu \right| \widehat{H}_{eff}\left| \mu \right\rangle
_{\mp }=\hbar \xi \mu /2.  \label{offdiagnew}
\end{equation}
When we assume that both $\delta $ and $\xi $ are small compared with the
splitting due to the interparticle interaction $\kappa $, so that we remain
in the Bragg regime, the two states $\left| \pm \mu \right\rangle $ remain
decoupled from the other number states, and we have an effective two-state
system. In practice, the parameter $\xi $ can be easily controlled, so that
many effects of two-state atoms \cite{Allen} can also realized for these two
states. For example, in analogy to the excitation of ground-state by an
adiabatic sweep across the resonance, one could create an effective transfer
from the state $\left| \mu \right\rangle $ to the state $\left| -\mu
\right\rangle $ by varying the parameter $\xi $ adiabatically from a
positive to a negative value that is large compared to $\Omega _{\mu }$.
This gives an effective collective transfer of $n=2\mu $ atoms from one well
to the other one.

\section{Time-dependent coupling}

When the coupling $\delta (t)$ varies with time, the time-dependent
eigenstates of the Hamiltonian are coupled to each other. The eigenstate
that correlates in the limit $\delta \rightarrow 0$ to the state $\left| \mu
\right\rangle _{\pm }$ is denoted as $\left| \varphi _{\mu }^{\pm
}\right\rangle $. Note that even eigenstates are only coupled to other even
eigenstates, and odd eigenstates to odd eigenstates. The coupling results
from the time dependence of the eigenstates. In fact, the term in the
Schr\={o}dinger equation coupling $\left| \varphi _{\mu }^{\pm
}\right\rangle $ to $\left| \varphi _{\nu }^{\pm }\right\rangle $ is
proportional to 
\begin{equation}
\left\langle \varphi _{\nu }^{\pm \text{ }}\left( t\right) \right| \frac{d}{%
dt}\left| \varphi _{\mu }^{\pm }(t)\right\rangle =-\left\langle \varphi
_{\nu }^{\pm \text{ }}\left( t\right) \right| \widehat{L}_{x}\left| \varphi
_{\mu }^{\pm \text{ }}\left( t\right) \right\rangle \frac{\hbar \stackrel{.}{%
\delta }\left( t\right) }{E_{\nu }^{\pm }-E_{\mu }^{\pm }};\text{ }\mu \neq
\nu .  \label{Adiabat1}
\end{equation}
This coupling is ineffective in the case that the r.h.s. of eq. (\ref
{Adiabat1}) is small compared with $(E_{\nu }^{\pm }-E_{\mu }^{\pm })/\hbar $%
. In this case, an initial eigenstate remains an eigenstate at all times.
This is the standard case of adiabatic following, which has been discussed
in the diffraction case \cite{Keller}. Since within the even or the odd
subspace there are no degeneracies, the dynamics of adiabatic following is
particularly simple. When the coupling coefficient $\delta $ is smoothly
switched on, with the system initially in the state $\left| \mu
\right\rangle =(\left| \mu \right\rangle _{+}+\left| \mu \right\rangle _{-})/%
\sqrt{2}$, the time-dependent state is obviously 
\begin{equation}
\left| \Psi (t)\right\rangle =e^{-i\vartheta (t)}\left( \left| \varphi _{\mu
}^{+}\right\rangle e^{-i\eta (t)/2}+\left| \varphi _{\mu }^{-}\right\rangle
e^{i\eta (t)/2}\right) /\sqrt{2},  \label{adiabat}
\end{equation}
with $\vartheta (t)=\int^{t}dt^{\prime }\left( E_{\mu }^{+}(t^{\prime
})+E_{\mu }^{-}(t^{\prime })\right) /2\hbar $ the average phase, and $\eta
(t)=\int^{t}dt^{\prime }\left( E_{\mu }^{+}(t^{\prime })-E_{\mu
}^{-}(t^{\prime })\right) /\hbar $ the accumulated phase difference of the
two eigenstates. In a time interval that the coupling $\delta $ is constant,
the phase difference $\eta (t)$ increases linearly with time, and the state (%
\ref{adiabat}) gives rise to expectation values oscillating at the single
frequency $\left( E_{\mu }^{+}(t^{\prime })-E_{\mu }^{-}(t^{\prime })\right)
/\hbar $. When the coupling is switched off again, the phase difference
approaches a constant limiting value $\overline{\eta }=\eta \left( \infty
\right) $. The state (\ref{adiabat}) at later times corresponds to a linear
superposition of the states $\left| \pm \mu \right\rangle $ proportional to $%
\left| \mu \right\rangle \cos (\overline{\eta }/2)+\left| -\mu \right\rangle
\sin (\overline{\eta }/2)$. Again, as we see, this effect that is known in
the diffraction case also has a counterpart for the double-well problem,
where adiabatic switching of the coupling between the wells leads to a
linear superposition of the Fock states $\left| n_{1},n_{2}\right\rangle
=\left| N/2+\mu ,N/2-\mu \right\rangle $ and $\left|
n_{1},n_{2}\right\rangle =\left| N/2-\mu ,N/2+\mu \right\rangle $. By proper
tailoring of the pulse, the final state can be made to coincide with either
one of these Fock states, or with the even state $\left| \mu \right\rangle
_{+}$ or with the odd state $\left| \mu \right\rangle _{-}$.

In contrast, when the coupling term $\delta (t)$ has the form of a short
pulse around time $0$, such that the action of the quadratic term can be
neglected during the pulse, the initial state $\left| \mu \right\rangle $
couples to all other states $\left| \mu ^{\prime }\right\rangle .$ The state
vector has exactly the same form (\ref{timedep}) as for diffraction in the
Raman-Nath regime. For the two-well problem, the evolution operator takes
the form $\widehat{U}=\exp (i\phi \widehat{L}_{x})$, with $\phi =\int
dt\delta (t)$, which has matrix elements that can be expressed in the Wigner
rotation matrices \cite{Edmonds} by 
\begin{equation}
\left\langle \mu ^{\prime }\right| \widehat{U}\left| \mu \right\rangle
=i^{\mu ^{\prime }-\mu }d_{\mu ^{\prime }\mu }^{J}(\phi ),  \label{wigner}
\end{equation}
with $J=N/2$. A comparison with Eq. (\ref{evol}) shows that for the
two-well-problem, the Wigner functions play the same role as the Bessel
functions in the diffraction case.

In Fig. 3 we plot the energy difference $E_{\mu }^{+}-E_{\mu }^{-}$ between
the even and odd eigenstate in the two-well case, for $N=100$, and for a few
values of $\delta /\kappa $. The splittings decrease monotonously for
increasing quantum number $\mu $.

\section{Conclusion}

In this paper we have analyzed both the similarity and the difference
between the dynamical behavior of atom diffraction from a standing wave and
a Bose-Einstein condensate in a double-well potential. In both cases, the
Hamiltonian is given by the generic form (\ref{lattice10}), the only
difference being in the commutation rules for the operators $\widehat{L}_{i}$%
. with $i=x,y,z$. Well-known diffraction phenomena as {\it Pendell\H{o}sung}
oscillations between opposite momenta in the case of Bragg diffraction, and
the result of adiabatic transitions between momentum states have
counterparts in the behavior of the atom distribution over the two wells, in
the case that the coupling between the wells is weak compared to the
interatomic interaction or slowly varying with time. A common underlying
reason for these effects is the symmetry of the Hamiltonian for inversion $%
\mu \leftrightarrow -\mu $, and the energy splitting between even and odd
states arising from the coupling term. In these cases, effective coupling
occurs between the states $\left| n_{1},n_{2}\right\rangle $ and $\left|
n_{2},n_{1}\right\rangle $ with opposite imbalance between the particle
numbers in the two wells. These states are coupled without population of the
intermediate states, so that a number of $n_{1}-n_{2}$ particles oscillate
collectively between the two wells. The interparticle interaction is
essential for this effect to occur. A simple analytical expression is
obtained for the {\it Pendell\H{o}sung} frequency. An initial state $\left|
n_{1},n_{2}\right\rangle $ with a well-determined number of atoms in each
well can be transferred to a linear superposition of $\left|
n_{1},n_{2}\right\rangle $ and $\left| n_{2},n_{1}\right\rangle $, which is
a highly entangled state of the two wells. A similar analogy is obtained to
diffraction in the Raman-Nath regime. For the double-well problem this
requires that the coupling is sufficiently short to ignore dynamical effect
of the atomic interaction during the coupling. The well-known diffraction
pattern in terms of the Bessel function is replaced by elements of theWigner
rotation matrix for the double well. These effects do not show up in the
mean-field approximation, where the Gross-Pitaevski equation holds.

\acknowledgments
This work is part of the research program of the ``Stichting voor
Fundamenteel Onderzoek der Materie'' (FOM).

\begin{figure}
\centering
\epsfig{file=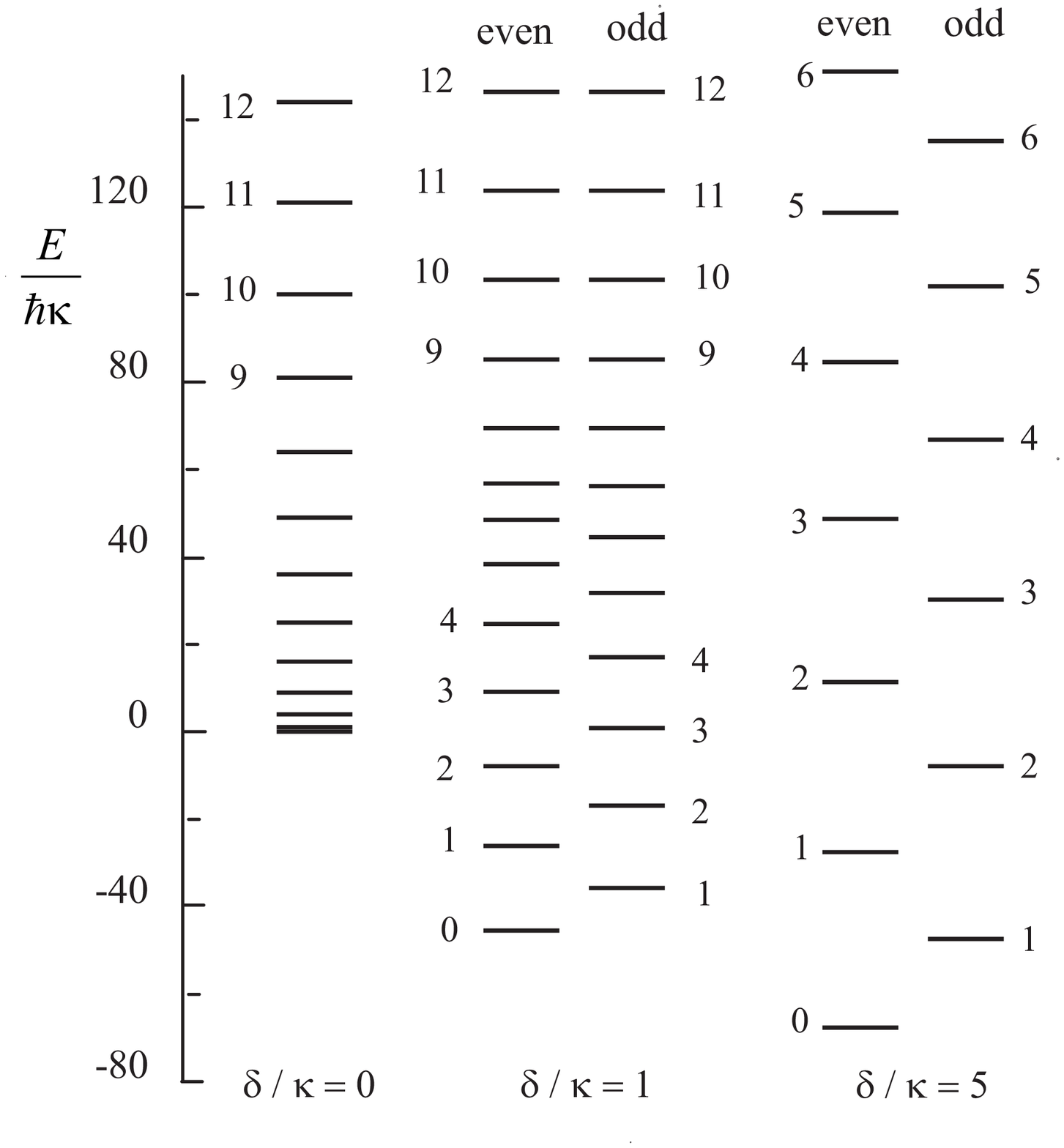,scale=0.5}
\caption{Energy levels in units of $\hbar \kappa $ for the double well with $%
N=100$ particles, for various values of $%
{\displaystyle {\delta  \over \kappa }}%
$. The levels are labeled by the quantum number $\mu $.}
\end{figure}

\begin{figure}
\centering
\epsfig{file=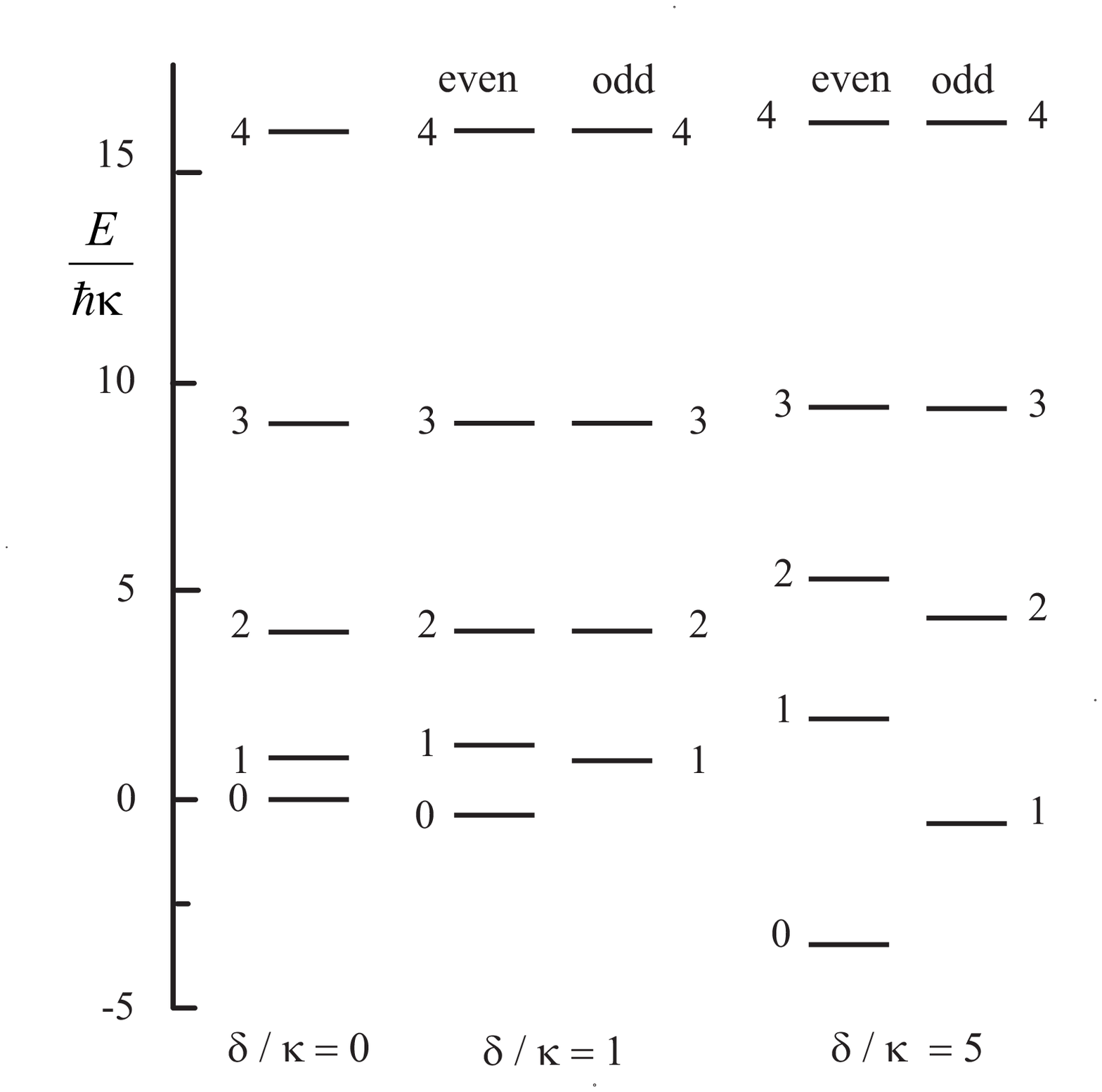,scale=0.5}
\caption{Energy levels in units of $\hbar \kappa $ for the diffraction case,
for various values of $%
{\displaystyle {\delta  \over \kappa }}%
$.}
\end{figure}

\begin{figure}
\centering
\epsfig{file=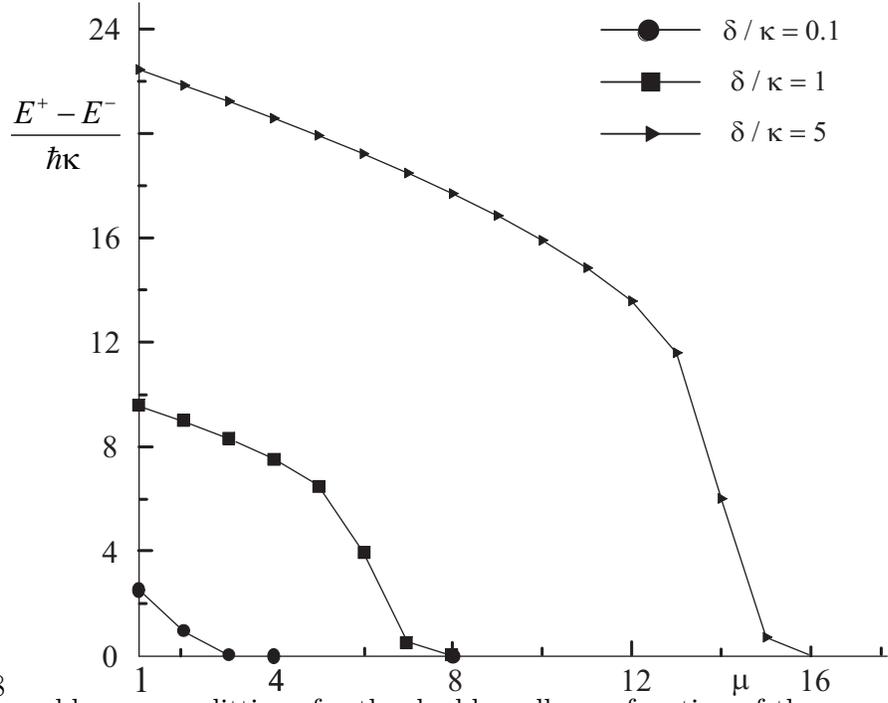,scale=0.8}
\caption{Even-odd energy splittings for the double well as a function of the
quantum number $\mu $, for various values of $%
{\displaystyle {\delta  \over \kappa }}%
$.}
\end{figure}

\end{document}